\documentclass[twocolumn]{jpsj2}
\usepackage{graphicx}
\usepackage{amsmath,amssymb}
\usepackage{color}

\def\runauthor{Youichi {\sc Yanase}}

\title{ % Don't Edit
Magneto-Electric Effect in Three-dimensional Coupled Zigzag Chains
} % Don't Edit

\author{Youichi {\sc YANASE}\footnote{E-mail:
yanase@phys.sc.niigata-u.ac.jp}}

\inst{Department of Physics, Niigata University, Niigata 950-2181, Japan 
}

\recdate{June 13, 2013}

\abst
{Stimulated by recent studies of quantum phases with broken local inversion symmetry, 
we study the magnetoelectric effect in locally noncentrosymmetric metals. 
We consider three-dimensional (3D) coupled zigzag chains and demonstrate that the antiferromagnetic moment 
is induced by the electric current through a staggered antisymmetric spin-orbit coupling. 
This current-induced magnetism is much larger than that in globally noncentrosymmetric metals. 
We provide an intuitive understanding of the current-induced antiferromagnetic moment by 
showing the inverse magnetoelectric effect, that is, the ferroic $p$-wave charge 
nematic order accompanied by the asymmetric band structure in the antiferromagnetic state.
We also examine conduction electrons coupled to localized spins via Kondo exchange coupling 
and demonstrate a significant enhancement of the magnetoelectric effect.
A possible experimental observation of the magnetoelectric effect in metals is discussed, 
with focus on LnM$_2$Al$_{10}$ compounds, such as NdRu$_2$Al$_{10}$ and TbRu$_2$Al$_{10}$. 
}

\kword
{
magnetoelectric effect, local inversion symmetry breaking, staggered Rashba spin-orbit coupling, 
current-induced magnetism, Kondo system}

\begin{document}
\sloppy
\maketitle

\newcommand{\vs}{\vspace*{3mm}}
\newcommand{\hs}{\noindent\hspace*{-3mm}}

\renewcommand{\k}{{\bf k}}
\newcommand{\K}{{\bf K}}
\renewcommand{\r}{\vec{r}}
\newcommand{\dd}{\mbox{\boldmath$d$}}
\newcommand{\kk}{\vec{k}'}
\newcommand{\kp}{\vec{k}_{+}}
\newcommand{\kkk}{\vec{k}''}
\newcommand{\q}{\vec{q}}
\newcommand{\Q}{\vec{Q}}
\newcommand{\qp}{\vec{q}_{+}}
\newcommand{\e}{\varepsilon}
\newcommand{\ee}{e}
\newcommand{\s}{{\mit{\it \Sigma}}}
\newcommand{\J}{\mbox{\boldmath$J$}}
\newcommand{\vv}{\mbox{\boldmath$v$}}
\newcommand{\Jh}{J_{{\rm H}}}
\newcommand{\LL}{\mbox{\boldmath$L$}}
\renewcommand{\SS}{\mbox{\boldmath$S$}}
\newcommand{\sSS}{\mbox{\boldmath$s$}}
\newcommand{\MM}{\mbox{\boldmath$M$}}
\newcommand{\g}{\mbox{\boldmath$g$}}
\newcommand{\Pauli}{\mbox{\boldmath$\sigma$}}
\newcommand{\Tc}{$T_{\rm c}$ }
\newcommand{\Tcf}{$T_{\rm c}$}
\newcommand{\Hc}{$H_{\rm c2}$ }
\newcommand{\Hcf}{$H_{\rm c2}$}
\newcommand{\etal}{{\it et al.}: }
\newcommand{\SRO}{Sr$_2$RuO$_4$ }
\newcommand{\SROf}{Sr$_2$RuO$_4$}
\newcommand{\kx}{k_{\rm c}}
\newcommand{\ky}{k_{\rm a}}
\newcommand{\kz}{k_{\rm b}}
\newcommand{\Neel}{N${\rm \acute{e}}$el } 
\newcommand{\Ca}{\mbox{\boldmath$a$}}
\newcommand{\Cn}{\mbox{\boldmath$n$}}

\section{Introduction}

Recent intensive research has clarified the intriguing roles of spin-orbit coupling 
in electron systems lacking the inversion symmetry. 
For instance, studies of noncentrosymmetric superconductivity~\cite{NCSC}, 
chiral magnetism~\cite{Chiral_Magnetism_1,Chiral_Magnetism_2}, 
multiferroics,~\cite{Katsura} spintronics such as the spin-Hall effect,~\cite{Spin_Hall_1, Spin_Hall_2} 
and topological quantum phases~\cite{RevModPhys.82.3045,RevModPhys.83.1057,Tanaka_review} 
have added new concepts in condensed matter physics. 

Although many previous works investigated metals, semiconductors, and insulators 
lacking the {\it global} inversion symmetry, recent advances in superconductivity 
have also elucidated novel quantum phases in {\it locally} noncentrosymmetric 
systems,~\cite{PhysRevB.84.184533,JPSJ.81.034702,PhysRevB.86.134514,
Yoshida-Sigrist-Yanase,Goryo2012} where the global inversion symmetry is not broken but 
atoms are not on the inversion center. 
Although a uniform antisymmetric spin-orbit coupling, such as Rashba spin-orbit coupling, 
plays a major role in globally noncentrosymmetric systems, a staggered antisymmetric 
spin-orbit coupling entangles the spin and orbital motions of electrons in locally 
noncentrosymmetric systems. 
Note that the staggered antisymmetric spin-orbit coupling is not smeared out by the global 
inversion symmetry, since the antisymmetric spin-orbit coupling is derived from the local 
properties of electrons, namely, local parity mixing and atomic 
LS coupling.~\cite{YanaseCePt3Si,Nagano-Shishidou-Oguchi} 
Indeed, its effect has been identified in recent experiments on the superlattice 
CeCoIn$_5$/YbCoIn$_5$~\cite{Goh}.

Motivated by the above advances, we investigate the magnetoelectric effect in locally 
noncentrosymmetric metals in this study. Spin-orbit coupling gives rise to many 
nonequilibrium magnetoelectric effects entangling electron spin and orbital motions; 
a typical one is the spin polarization induced by electric current~\cite{Ivchenko,Edelstein,Springer}, 
which is closely related to the spin-Hall effect~\cite{Culcer,Gorini}. 
Although the current-induced spin-polarization in noncentrosymmetric systems was predicted 
more than three decades ago~\cite{Ivchenko,Edelstein,Springer} and observed 
in semiconductors~\cite{Y.K.Kato,Silov,Sih,Yang,Stern}, 
experimental observation has not been successful for metals partly because the 
induced magnetic moment is small as $\sim 10^{-10} \mu_{\rm B}$ per unit cell. 
We show that a rather large antiferromagnetic moment can be induced by electric current 
in locally noncentrosymmetric metals. 
As naturally expected, this magnetoelectric effect will be enhanced further 
by Kondo exchange coupling with localized spins.  
We estimate the order of the current-induced antiferromagnetic moment and propose an experimental study 
of $f$-electron systems such as LnM$_2$Al$_{10}$ compounds (Ln=lanthanoid ions, M=transition metal ions).

In this research, we study 3D coupled zigzag chains, which are a typical crystal structure accompanied 
by the local violation of the inversion symmetry. LnM$_2$Al$_{10}$ compounds as well as 
ferromagnetic heavy-fermion superconductors UCoGe, URhGe, and UGe$_2$~\cite{Aoki_review} have 
such crystal structure. 
In Sect.~2, we introduce a model of conduction electrons affected by a staggered antisymmetric 
spin-orbit coupling. 
The current-induced antiferromagnetic moment is formulated in Sect.~3.1. 
The ferroic $p$-wave charge nematic order and asymmetric band structure induced in the 
antiferromagnetic state are shown in Sect.~3.2, and enable an intuitive understanding of 
the magnetoelectric effect.  
Numerical results are obtained for 1D zigzag chains and 3D coupled zigzag chains in Sects.~4 and 5, 
respectively. An enhancement of the magnetoelectric effect in Kondo systems is discussed in Sect.~6. 
A brief summary and several discussions are given in Sect.~7.

\section{Model of 3D Coupled Zigzag Chains}

\begin{figure}[htbp]
\begin{center}
\includegraphics[width=7cm,angle=0]{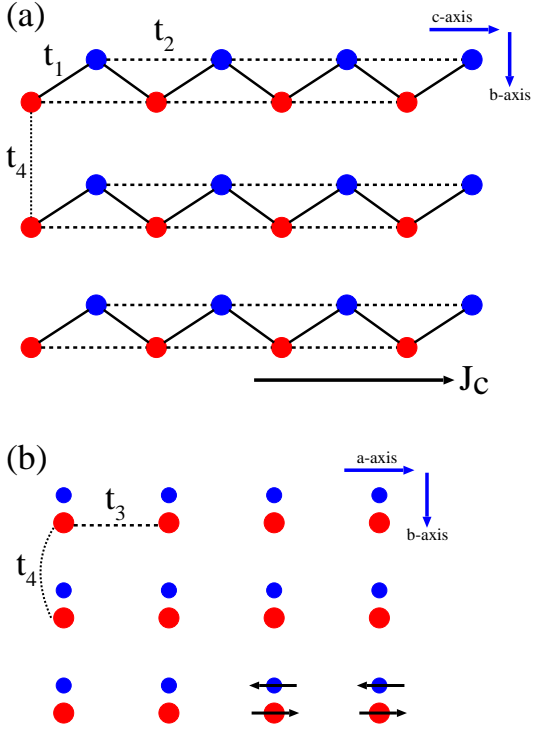}
\caption{(Color online) 
Crystal structure of 3D coupled zigzag chains. 
(a) Projection along the {\it a}-axis. (b) Projection along the {\it c}-axis. 
Dark (blue) and light (red) circles depict the $a$- and $b$-sublattices, respectively. 
The hopping integrals are shown as $t_1 - t_4$. 
The long arrow in Fig.~1(a) shows the direction of electric current, $J_{\rm c}$. 
Arrows in Fig.~1(b) show the current-induced ``antiferromagnetic moment''. 
}
\end{center}
\end{figure}

First, we introduce a model describing 3D coupled zigzag chains, 
\begin{eqnarray}
\label{H_0}
&& \hspace*{-8mm}  
H = \sum_{\tiny{\k} s} 
\e(\k) \left[a_{\k s}^{\dag} b_{\k s} + {\rm c.c.}\right] 
\nonumber \\ && \hspace*{-2mm}
+  \sum_{\k s}  \e'(\k) \left[a_{\k s}^{\dag} a_{\k s} + b_{\k s}^{\dag} b_{\k s} \right] 
\nonumber \\ && \hspace*{-2mm}
+ \alpha \sum_{\k s s'} \g(\k) \Pauli_{ss'} \left[a_{\k s}^{\dag} a_{\k s'} - b_{\k s}^{\dag} b_{\k s'} \right],  
\end{eqnarray}
where $a_{\k s}$  and $b_{\k s}$ are the annihilation operators of electrons 
with spin $s = \uparrow, \downarrow$ on the sublattices $a$ and $b$, respectively. 
The single electron kinetic energy terms $\e(\k)$ and $\e'(\k)$  
are described by taking into account the nearest- and next-nearest-neighbour hoppings 
in a one-dimensional (1D) zigzag chain as well as the nearest-neighbour interchain hopping, 
\begin{eqnarray}
&&  \hspace*{-12mm}
\e(\k)  = -2 t_1 \cos \frac{\kx}{2},  
%- 2 t_3 \cos \frac{3 \kx}{2}, 
\\ &&  \hspace*{-12mm}
\e'(\k)  = -2 t_2 \cos \kx - 2 t_3 \cos \ky - 2t_4 \cos \kz.  
\end{eqnarray}
The crystal structure of 3D coupled zigzag chains and the hopping integrals $t_1$-$t_4$  
are illustrated in Fig.~1. 

We are focused on the violation of local inversion symmetry which induces 
the staggered antisymmetric spin-orbit coupling, the last term of Eq.~(\ref{H_0}). 
Since we consider quasi-1D systems, the g-vector is approximated as $\g(\k) = \sin \kx \hat{z} $.  
We choose the crystallographic {\it a}-axis [see Fig.~1] as the quantization axis of the spin, 
namely, $\hat{z} = \hat{a}$.

We here describe the model in matrix form using the vector operator 
$\hat{C}^{\dag}_{\k} = 
\left(a^{\dag}_{\k \uparrow}, 
b^{\dag}_{\k \uparrow},
b^{\dag}_{\k \downarrow}, 
a^{\dag}_{\k \downarrow}\right)$: 
\begin{eqnarray}
H= \sum_{\k} \hat{C}^{\dag}_{\k} \hat{H}_4(\k) \hat{C}_{\k}. 
\label{H_k}
\end{eqnarray}
Note that the four components in the vector operator are aligned so as to simplify 
the following description. 
Because the $z$-component of the spin is conserved in this model, 
the $4 \times 4$ matrix $ \hat{H}_4(\k) $ is block-diagonalized as 
%
%%%\twocolumn[
%
\begin{eqnarray}
\label{4x4}
&& \hspace{-10mm} \hat{H}_4(\k) = 
%\nonumber \\ && \hspace*{-20mm}
\left(
\begin{array}{cc}
\hat{H}_2(\k) & \hat{0}
\\
\hat{0} & \hat{H}_2(\k)
\\
\end{array}
\right),  
%\nonumber \\ &&
%
\end{eqnarray}
where
\begin{eqnarray}
\label{2x2}
&& \hspace{-10mm} \hat{H}_2(\k) = 
%\nonumber \\ && \hspace*{-20mm}
\left(
\begin{array}{cc}
\e'(\k) + \alpha \sin \kx & \e(\k)
\\
\e(\k) & \e'(\k) - \alpha \sin \kx
\\
\end{array}
\right). 
%\nonumber \\ &&
%
\end{eqnarray}
%
%%%]
%
Diagonalizing this matrix, we obtain the dispersion relation  
\begin{eqnarray}
&& \hspace{-10mm} 
E_\pm(\k) = \e'(\k) \pm \sqrt{\e(\k)^2 + \alpha^2 \sin^2 \kx}. 
\end{eqnarray}

In order to understand the above electronic structure, 
we here consider two extreme parameters. 
In the absence of inter-sublattice electron hopping, namely, $t_1 =0$, the dispersion relation is reduced 
to $E_\pm(\k) = \e'(\k) \pm \alpha \sin \kx $. This band structure reproduces that of 
noncentrosymmetric metals~\cite{Springer}, but there remains a twofold degeneracy in our case, 
as guaranteed by the global inversion symmetry and time reversal symmetry.  
For instance, the electron state with the spin $s = \uparrow$ on the $a$-sublattice is degenerate 
with that having the opposite spin $s = \downarrow$ on the $b$-sublattice. Thus, the electron spin 
is entangled with the sublattice degree of freedom. On the other hand, 
when the inter-sublattice hopping is much larger than the spin-orbit coupling as $t_1 \gg \alpha$, 
we obtain the conventional band structure consisting of the bonding and anti-bonding orbitals. 
In this case, the twofold degeneracy arises from the spin degree of freedom in a conventional way. 
We will see the crossover between these electronic states in Figs.~3 and 6.

\section{Magnetoelectric Effect}

\subsection{Current-induced antiferromagnetic moment}

We here investigate the magnetoelectric effect in locally noncentrosymmetric conducting electrons. 
In contrast to the current-induced spin polarization in noncentrosymmetric 
systems~\cite{Ivchenko,Edelstein,Springer}, an ``antiferromagnetic moment'' is induced 
when the electric field is applied to locally noncentrosymmetric metals. 
Strictly speaking, the spin is antiferromagnetically aligned {\it in the unit cell}, 
as shown in Fig.~1(b). The wave number is $\q =0$ since the translation symmetry is not broken. 
This magnetic structure is regarded as a linear combination of the toroidal magnetic moment and 
magnetic quadrupole moment.~\cite{Kusunose-Arima}

In a linear-response region, the magnetoelectric effect is described as 
\begin{eqnarray}
&& \hspace{-20mm} M^{\rm AF}_{\mu} = - \Upsilon_{\mu\nu} E_{\nu}, 
\end{eqnarray}
where $M^{\rm AF}_{\mu}$ is the antiferromagnetic moment and $E_{\nu}$ is the electric field.  
The magnetoelectric coefficient $\Upsilon_{\mu\nu}$ is calculated using the 
standard Kubo formula  
\begin{eqnarray}
&& \hspace{-10mm}
\Upsilon_{\mu\nu} = {\rm lim}_{\omega \rightarrow 0} \frac{1}{i \omega} 
K_{\mu\nu}^{\rm ME} (i \omega_n) |_{i \omega_n \rightarrow \omega + i 0}. 
\end{eqnarray}
The response function is obtained as 
\begin{eqnarray}
&& \hspace{-14mm} 
K_{\mu\nu}^{\rm ME} (i \omega_n) = \frac{1}{2} g \mu_{\rm B} \int_{0}^{1/T} {\rm d}\tau 
\langle T_{\tau} {S_{\mu}^{\rm AF}(\tau)J_{\nu}(0)} \rangle e^{i \omega_n \tau}, 
\end{eqnarray}
where $S_{\mu}^{\rm AF}$ and $J_{\nu}$ are the operators of 
antiferromagnetic spin and electric current, respectively. 
They are described as  
\begin{eqnarray}
&& \hspace{-10mm} S_{\mu}^{\rm AF} = 
\sum_{\k} \hat{C}^{\dag}_{\k} \hat{\sigma}^{\rm AF} \hat{C}_{\k}, 
\\ && \hspace{-10mm}
J_{\nu} = e \sum_{\k} \hat{C}^{\dag}_{\k} \hat{v}_{\nu}(\k) \hat{C}_{\k}, 
\end{eqnarray}
using the matrices  
\begin{eqnarray}
&& \hspace{-10mm} \hat{\sigma}^{\rm AF} = 
%\nonumber \\ && \hspace*{-20mm}
\left(
\begin{array}{cc}
\hat{\sigma}_{\rm z} & \hat{0}  \\
\hat{0} &  \hat{\sigma}_{\rm z} \\
\end{array}
\right), 
%\left(
%\begin{array}{cccc}
%1 & 0 & 0 & 0 \\
%0 & -1 & 0 & 0 \\
%0 & 0 & 1 & 0 \\
%0 & 0 & 0 & -1 \\
%\end{array}
%\right), 
%\nonumber \\ &&
%
\end{eqnarray}
and 
%
%\begin{eqnarray}
%&& \hspace{-20mm} 
$
\hat{v}_{\nu}(\k) = \partial \hat{H}_{4}(\k)/\partial k_{\nu}. 
$
%\end{eqnarray}

The magnetoelectric coefficient $\Upsilon_{\rm zc}$ is finite, and others are zero 
in our model. 
Thus, the antiferromagnetic moment along the {\it a}-axis is induced by the electric current 
along the {\it c}-axis, as illustrated in Fig.~1. 
Another magnetoelectric coefficient, $\Upsilon_{\rm xa}$, is allowed by symmetry, but 
it disappears when we assume a 1D g-vector, $\g(\k) = \sin \kx \hat{z} $. 

Using Green functions, the magnetoelectric coefficient is described as 
\begin{eqnarray}
&& \hspace{-8mm} 
\Upsilon_{\rm zc} = -g \mu_{\rm B} |e| 
%\nonumber \\ && \hspace{-8mm}
% \times
\sum_{\k} 
\frac{\alpha \sin \kx}{\sqrt{\e(\k)^2 + \alpha^2 \sin^2 \kx }} 
\nonumber \\ && \hspace{-0mm}
\times 
\left[v_{\rm c}^{(+)}(\k) K_{+}(\k) - v_{\rm c}^{(-)}(\k) K_{-}(\k)  \right],
%\nonumber \\ && \hspace{-8mm}
\end{eqnarray}
where 
%
%\begin{eqnarray}
%&& \hspace{-10mm} 
$
v_{\rm c}^{(\pm)}(\k) =  \partial E_\pm(\k)/\partial \kx, 
$
%\end{eqnarray}
%
and 
\begin{eqnarray}
&& \hspace{-10mm} 
K_\pm(\k) = \frac{1}{\pi} \int_{-\infty}^{\infty} {\rm d}\e \left(-f'(\e) \right) 
\left[ {\rm Im} G_\pm^{\rm R} (\k,\e)\right]^2. 
\end{eqnarray}
Taking into account the finite lifetime of quasiparticles, $\tau_\pm(\k)$, 
the retarded Green function in the band basis is described as 
$G_\pm^{\rm R} (\k,\e)= \left(\e -E_\pm(\k) + i/2\tau_\pm(\k) \right)^{-1}$. 
Thus, we obtain
\begin{eqnarray}
&& \hspace{-10mm} 
K_\pm(\k) = \tau_\pm(\k) \left[-f'(E_\pm(\k)) \right],  
\end{eqnarray}
where $f'(\epsilon)$ is the derivative of the Fermi distribution function. 
In this study, we do not discuss the source of quasiparticle scattering and 
assume the momentum- and band-independent lifetime $\tau = \tau_\pm(\k)$, 
for simplicity. Although the spin-Hall effect significantly depends on the scatterer~\cite{Inoue}, 
the magnetoelectric effect is not sensitive to it~\cite{Edelstein}. 
Thus, the magnetoelectric coefficient is obtained as  
\begin{eqnarray}
&& \hspace{-7mm} 
\Upsilon_{\rm zc} = - g \mu_{\rm B} |e| \tau 
\sum_{\k} \frac{\alpha \sin \kx}{\sqrt{\e(\k)^2 + \alpha^2 \sin^2 \kx }} 
\nonumber \\ && \hspace{-3mm}
%\times \sigma v_{\rm c}^{(\sigma)}(\k) \left[-f'(E_\sigma(\k)) \right].  
\times \left\{v_{\rm c}^{(+)}(\k) \left[-f'(E_+(\k)) \right] - v_{\rm c}^{(-)}(\k) \left[-f'(E_-(\k)) \right]  
\right\}. 
\nonumber \\ && \hspace{-8mm}
\label{Upsilon}
\end{eqnarray}
We clearly see that the magnetoelectric effect vanishes 
in the absence of the staggered antisymmetric spin-orbit coupling $\alpha$. 

On the basis of the same assumptions as above, the electric conductivity along the {\it c}-axis is 
obtained as 
\begin{eqnarray}
\label{sigma}
&& \hspace{-10mm}
\sigma_{\rm c} = 2 e^2 \tau \sum_{\k} 
\sum_{\sigma = \pm} v_{\rm c}^{(\sigma)}(\k)^2 \left[-f'(E_\sigma(\k)) \right]. 
%\nonumber \\ && \hspace{-5mm}
%\left\{v_{\rm c}^{(1)}(\k)^2 \left[-f'(E_1(\k)) \right] 
%+ v_{\rm c}^{(2)}(\k)^2 \left[-f'(E_2(\k)) \right] \right\}. 
%\nonumber \\ && \hspace{-5mm}
\end{eqnarray}
Since the ratio $\Upsilon_{\rm zc}/\sigma_{\rm c}$ is independent of the phenomenological parameter $\tau$, 
we describe the magnetoelectric effect with the use of the electric current  
$J_{\rm c} =\sigma_{\rm c} E_{\rm c}$ as   
\begin{eqnarray}
&& \hspace{-10mm} M^{\rm AF}_{\rm z} = - \frac{\Upsilon_{\rm zc}}{\sigma_{\rm c}} \sigma_{\rm c}  E_{\rm c}
%\\ && \hspace{-30mm}
=- \frac{g\mu_{\rm B}}{2|e|D} \Gamma_{\rm zc} J_{\rm c}, 
\label{Gamma}
\end{eqnarray}
where $D$ is the band width and $\Gamma_{\rm zc}$ is the dimensionless magnetoelectric coefficient (DMEC). 
We numerically calculate the DMEC in Sects.~4 and 5.

\subsection{Asymmetric electronic structure in the antiferromagnetic state}

We here study the inverse magnetoelectric effect in locally noncentrosymmetric metals. 
We consider the ``antiferromagnetic state'' where the magnetic moment is spontaneously ordered, 
as in Fig.~1(b). 
The electronic structure is calculated by taking the molecular field of 
antiferromagnetic order, $H_{\rm MF} = - h^{\rm AF} S_{\rm z}^{\rm AF}$, into account.  
The single-particle Hamiltonian $H_{\rm AFM} = H + H_{\rm MF}$ is block-diagonalized as in Eq.~(\ref{4x4}), 
and the $2 \times 2$ matrix is obtained as 
\begin{eqnarray}
\label{2x2AF}
&& \hspace{-10mm} \hat{H}_2(\k) = 
\nonumber \\ && \hspace*{-10mm}
\left(
\begin{array}{cc}
\e'(\k) + \alpha \sin \kx - h^{\rm AF} & \e(\k)
\\
\e(\k) & \e'(\k) - \alpha \sin \kx + h^{\rm AF}
\\
\end{array}
\right). 
\nonumber \\ &&
\end{eqnarray}
Thus, we obtain the asymmetric band structure 
\begin{eqnarray}
\label{dispersionAF}
&& \hspace{-15mm} 
E_\pm(\k) = \e'(\k) \pm \sqrt{\e(\k)^2 + (\alpha \sin \kx - h^{\rm AF})^2}, 
\end{eqnarray}
in which $E_\pm(\k) \ne E_\pm(-\k)$. 
An example of the band structure is drawn in Fig.~2(a).
We see that the antiferromagnetic moment polarizes the ``momentum'' of conducting electrons 
through spin-orbit coupling. 
\begin{figure}[htbp]
\begin{center}
\includegraphics[width=8cm,angle=0]{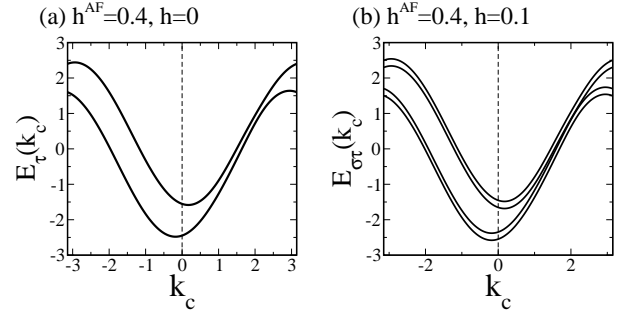}
\caption{
Asymmetric band structure of zigzag chains in the ``antiferromagnetic state''. 
(a) Two asymmetric bands with twofold degeneracy at zero magnetic field, $h=0$ [Eq.~(\ref{dispersionAF})]. 
(b) Four non-degenerate asymmetric bands in the magnetic field along the $a$-axis, 
$h=0.1$ [Eq.~(\ref{dispersionAF+F})]. 
We assume the antiferromagnetic molecular field $h^{\rm AF} =0.4$ and 
choose $t_1 = 0.1$,  $t_2 = 1$, and $t_3 = t_4 = 0$. 
These unconventional band structures should be compared with the symmetric band structure 
for $h^{\rm AF} = h = 0$ [see Fig.~3(c)]. 
}
\end{center}
\end{figure}

This asymmetric band structure is understood by considering the induced order parameter due to 
spin-orbit coupling. 
As the staggered spin-singlet superconductivity (pair density wave state) induces 
the uniform spin-triplet superconductivity through the staggered antisymmetric spin-orbit 
coupling,~\cite{Yoshida_preprint} 
the ferroic $p$-wave charge nematic order is induced in the antiferromagnetic state of zigzag chains. 
A finite order parameter of the $p$-wave charge nematic order, 
$N_{\rm c}^{\rm p} = \sum_{k_{\rm c}} \langle \sin k_{\rm c} \, n(k_{\rm c}) \rangle$, indicates  
the asymmetric electric structure. In this way, 
the asymmetric band structure in Fig.~2(a) is attributed to the $p$-wave charge nematic order 
induced by spin-orbit coupling.

We also understand the asymmetric band structure from the viewpoint of symmetry. 
In the locally noncentrosymmetric crystal structure, the collinear antiferromagnetic order 
may break the inversion symmetry ($P$-symmetry) as well as the time-reversal symmetry 
($T$-symmetry).~\cite{Comment1}  
This is the case that we consider in this subsection. 
These broken symmetries lead to the asymmetric band structure, where the momentum $\k$ 
is not equivalent to $-\k$. 
Note that the combined $PT$-symmetry protects the twofold degeneracy 
at each momentum. 
Because the Hamiltonian $H_{\rm AFM}$ is invariant under the successive operations of 
time reversal and spatial inversion, by which the quantum numbers change as 
$(\k,\sigma) \rightarrow (-\k,-\sigma) \rightarrow (\k,-\sigma)$, there remains a twofold 
degeneracy in the band structure~\cite{Arima_discussion}. 

The $PT$-symmetry is broken by applying a magnetic field along the $a$-axis. 
Then, the twofold degeneracy is simply lifted as  
\begin{eqnarray}
\label{dispersionAF+F}
&& \hspace{-10mm} 
E_{\sigma\tau}(\k) = E_{\tau}(\k) + \sigma h, 
%\nonumber \\ &&
\end{eqnarray}
where $\tau = \pm 1$, $\sigma = \pm 1$, and $h$ is the Zeeman energy. 
For instance, we show the four non-degenerate asymmetric bands for $h=0.1$ [Fig.~2(b)]. 
The ferroelectric polarization is induced in this state; the $a$-sublattice and $b$-sublattice of 
zigzag chains have nonequivalent charge densities.

Note that the asymmetric band structure does not yield spontaneous electric current. 
In Appendix, we prove that the spontaneous 
electric current vanishes independent of the structure of the crystal lattice, 
magnetic order, and spin-orbit coupling. 
On the other hand, the asymmetric band structure itself can be measured in the experiments 
using de Haas-van Alphen oscillation measurements or angle-resolved photo-emission 
spectroscopy (ARPES).
A magnetic structure similar to that studied in this subsection has been clarified  
in LnM$_2$Al$_{10}$ compounds, which we will discuss in Sect.~6.

The current-induced antiferromagnetic moment formulated in Sect.~3.1 is an inverse effect of 
the asymmetric band structure induced by the antiferromagnetic order. 
The dissipative electric current deforms the Fermi surface and induces the antiferromagnetic moment. 
This is an intuitive understanding of the magnetoelectric effect in locally noncentrosymmetric metals.
An analogous explanation has been discussed for the current-induced spin polarization 
in noncentrosymmetric metals~\cite{Springer}.

\section{1D Zigzag Chain}

In the following sections, we calculate the current-induced antiferromagnetic moment 
on the basis of the formulation given in Sect.~3.1.  
In this section, we study 1D models where $t_3 = t_4 =0$, and elucidate 
the effects of the band structure on the DMEC, $\Gamma_{\rm zc}$.
Observations in this section will be the basis of our understanding of 
more realistic 3D coupled zigzag chains, as we will show in Sect.~5.

\subsection{Deep zigzag chain}

First, we investigate deep zigzag chains, whose crystal structure is illustrated in Fig.~3(a). 
In this case, we consider that the inter-sublattice hopping is smaller than the intra-sublattice one, 
namely, $t_1 \le t_2$. For $t_1 \ll t_2$, the band structure resembles that of noncentrosymmetric systems, 
as shown in Fig.~3(b). With increasing $t_1$, a gap opens at the $\Gamma$-point ($k_{\rm c}=0$), as 
demonstrated in Figs.~3(c) and 3(d).

We find that the magnetoelectric effect does not occur in purely 1D chains 
without a sublattice structure. 
Indeed, the magnetoelectric coefficient $\Upsilon_{\rm zc}$ vanishes at $t_1 = 0$ 
even in the presence of an antisymmetric spin-orbit coupling, 
because the two terms in Eq.~(\ref{Upsilon}) are completely canceled out 
for the band-independent quasiparticle lifetime. 
In other words, the magnetoelectric effect of zigzag chains is induced 
by the inter-sublattice hopping $t_1$.

\begin{figure}[htbp]
\begin{center}
\includegraphics[width=8cm,angle=0]{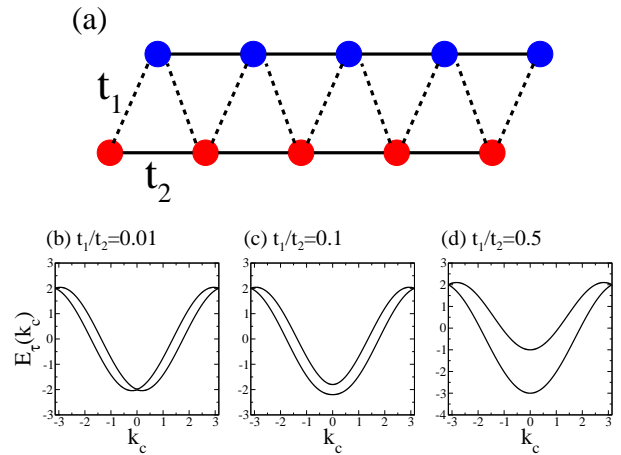}
\caption{(Color online) 
(a) Crystal structure of a deep 1D zigzag chain. 
Strongly (weakly) coupled bonds are shown by solid (dashed) lines. 
(b)-(d) Band structures for the inter-sublattice hopping, 
$t_1 = 0.01$, $0.1$, and $0.5$, respectively. 
We fix the intra-sublattice hopping to $t_2 =1$ and assume the spin-orbit coupling $\alpha=0.4$. 
}
\end{center}
\end{figure}

Figure 4 shows the DMEC as a function of the chemical potential. It is shown that the 
magnetoelectric effect is significantly enhanced around the lower band edge. 
When the inter-sublattice hopping $t_1$ is small, a large DMEC is realized in the narrow region 
of chemical potential. 

\begin{figure}[htbp]
\begin{center}
\includegraphics[width=7cm,angle=0]{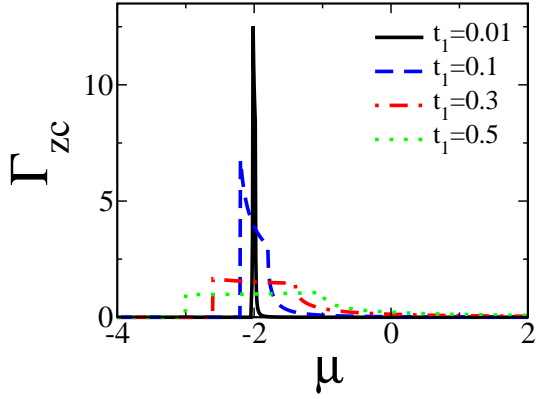}
\caption{(Color online) 
DMEC defined in Eq.~(\ref{Gamma}) as a function of the chemical potential. 
We show the results for $t_1 =0.01$, $0.1$, $0.3$, and $0.5$ at zero temperature.  
The other parameters are the same as those in Fig.~3.
}
\end{center}
\end{figure}

We understand this enhancement of the magnetoelectric effect in zigzag chains 
by simply looking at the band structure. 
As illustrated in Fig.~5, the $E_+$-band does not cross the Fermi level when the chemical potential is 
in the gap at the $\Gamma$-point, namely, $\mu_1 < \mu < \mu_2$ 
where  $\mu_1 = \e'(0) - |\e(0)|$ and $\mu_2 = \e'(0) + |\e(0)| $. 
In this case, the first term of Eq.~(\ref{Upsilon}) vanishes and the second term gives rise to a large  
DMEC exceeding unity, $\Gamma_{\rm zc} > 1$. 
On the other hand, the cancellation of the two terms in Eq.~(\ref{Upsilon}) leads to a small DMEC, 
$\Gamma_{\rm zc} \ll 1$, for a large chemical potential $\mu_2 < \mu < \mu_3 = \e'(\pi,\pi)$. 
When the chemical potential is around the upper band edge, $\mu_3 < \mu$, 
the contributions of four Fermi points of the $E_+$-band cancel each other out, giving rise to 
a small DMEC, $\Gamma_{\rm zc} \ll 1$. 
These cancellations of the magnetoelectric effect are unavoidable 
in noncentrosymmetric systems~\cite{Springer}, but can be avoided in locally 
noncentrosymmetric zigzag chains. 
This is the reason why a large magnetoelectric effect appears in the latter, 
as we demonstrated in Fig.~4.

\begin{figure}[htbp]
\begin{center}
\includegraphics[width=6cm,angle=0]{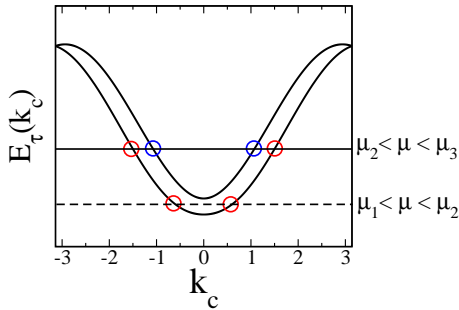}
\caption{(Color online) 
Band structure and Fermi points of 1D zigzag chains. 
For $\mu_1 < \mu < \mu_2$, the $E_-$-band has two Fermi points (red circles), while 
the $E_+$-band does not cross the Fermi level. 
For $\mu_2 < \mu < \mu_3$, both the $E_+$- and $E_-$-bands have two Fermi points. 
}
\end{center}
\end{figure}

\subsection{Shallow zigzag chain}

\begin{figure}[htbp]
\begin{center}
\includegraphics[width=8cm,angle=0]{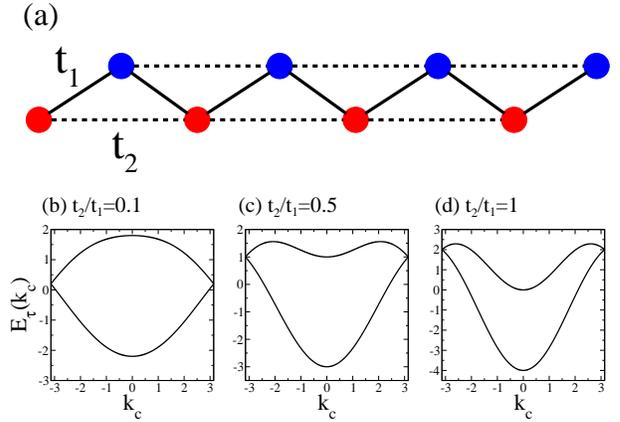}
\caption{(Color online) 
(a) Crystal structure of a shallow 1D zigzag chain. 
(b)-(d) Band structures for the intra-sublattice hopping 
$t_2 = 0.1$, $0.5$, and $1$, respectively. 
We fix the inter-sublattice hopping to $t_1 =1$ and assume the spin-orbit coupling $\alpha=0.4$.
}
\end{center}
\end{figure}

Next, we investigate the shallow 1D zigzag chains illustrated in Fig.~6(a). 
When the inter-sublattice hopping is much larger than the intra-sublattice one
[for instance, $t_2/t_1 =0.1$ in Fig.~6(b)], the cancellation of the two terms in Eq.~(\ref{Upsilon}) 
does not occur. 
Therefore, a moderate DMEC, $\Gamma_{\rm zc} \sim 1$, is obtained irrespective of the chemical potential, 
as shown in Fig.~7. The DMEC is particularly enhanced around the upper band edge, $\mu = 1.8$, 
because of the small Fermi velocity of the upper band around the $\Gamma$ point [$k_{\rm c} =0$]. 
Note that the DMEC is inversely proportional 
to the Fermi velocity, $\Gamma_{\rm zc} \propto 1/v_{\rm F}$. 
%as the electric conductivity is proportional to the Fermi velocity. 
With increasing $t_2/t_1$, the DMEC decreases around the upper band edge 
because of the cancellation of four Fermi points in the $E_+$-band.

\begin{figure}[htbp]
\begin{center}
\includegraphics[width=7cm,angle=0]{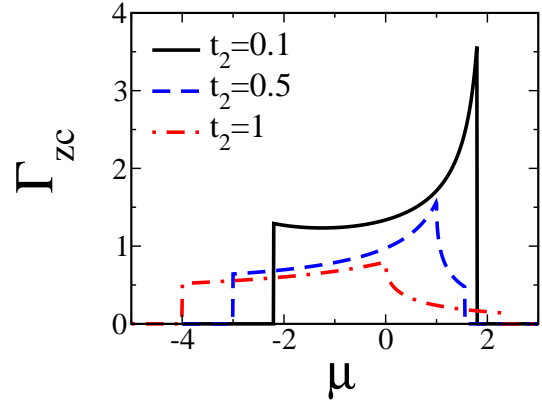}
\caption{(Color online) 
DMEC at zero temperature for the intra-sublattice hoppings $t_2 = 0.1$, $0.5$, and $1$.  
The other parameters are the same as those in Fig.~6. 
}
\end{center}
\end{figure}

Summarizing Sects.~4.1 and 4.2, a large magnetoelectric effect occurs in both deep and shallow 
zigzag chains when the cancellation of the two bands is suppressed by the band structure. 
The DMEC is particularly large in deep zigzag chains at low carrier densities. 
Thus, the current-induced antiferromagnetic moment in zigzag chains can be much larger than 
the current-induced magnetic moment in noncentrosymmetric metals.

\section{3D Coupled Zigzag Chains}

Now we turn to 3D coupled zigzag chains. 
Taking the isotropic inter-zigzag-chain couplings into account, $t_3 = t_4$, we clarify the 
effect of three-dimensionality on the magnetoelectric effect.
Figure~8 shows the DMEC as a function of the chemical potential and inter-zigzag-chain coupling.

\begin{figure}[htbp]
\begin{center}
\vspace{-7mm}
\includegraphics[width=6cm,angle=0]{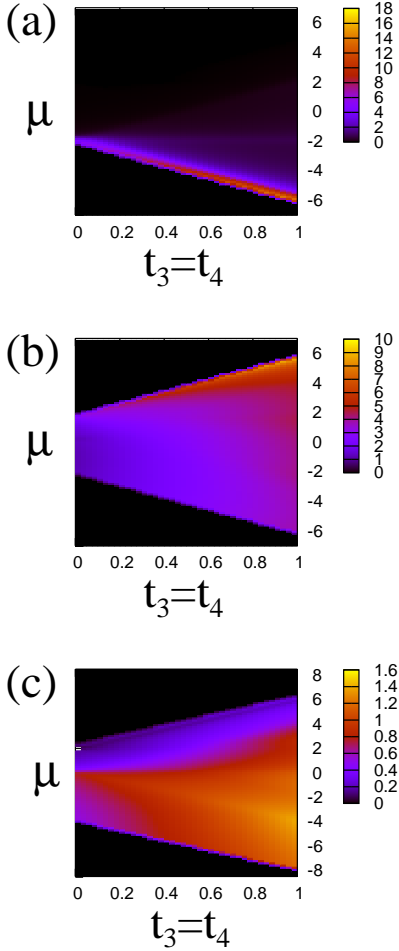}
\vspace{-5mm}
\caption{(Color online) 
DMEC as a function of the chemical potential $\mu$ and isotropic inter-zigzag-chain coupling $t_3 = t_4$, 
at a low temperature $T=0.02$. 
We assume $(t_1, t_2) =$ $(0.1, 1)$, $(1, 0.1)$, and $(1, 1)$ in (a), (b), and (c), respectively. 
}
\end{center}
\end{figure}

In the case of deep zigzag chains, $(t_1, t_2) = (1, 0.1)$, a pronounced magnetoelectric effect occurs,  
as shown by a large DMEC around the lower band edge in Fig.~8(a). 
The mechanism of the enhancement of the DMEC has been shown for 1D zigzag chain; 
the $E_+$-band does not cross the Fermi level when 
$\e'(0) - |\e(0)| < \mu < \e'(0) + |\e(0)|$.  
Furthermore, Fig.~8(a) shows an increase in the DMEC with increasing 3D coupling, $t_3 = t_4$. 

The three-dimensionality also enhances the magnetoelectric coupling of shallow zigzag chains. 
Figure~8(b) shows an increase in the DMEC with increasing 3D coupling, 
particularly around the upper band edge. 
Although an enhancement of the magnetoelectric effect due to 3D coupling appears 
in the intermediate case, $t_1 = t_2$, we obtain a small DMEC, as shown in Fig.~8(c). 
These results demonstrate that 
the three-dimensionality does not disturb the magnetoelectric effect in coupled zigzag chains.

Finally, we estimate the current-induced antiferromagnetic moment of conducting electrons. 
For a quantitative estimation, we denote the lattice constant along the $a$-, $b$-, and $c$-axes, 
as $a$ [cm], $b$ [cm], and $c$ [cm], respectively.  
We consider the macroscopic current density $j$ [A/cm$^2$], which leads to the current 
per zigzag chain, $J_{\rm c} = j ab$ [A/chain]. 
The induced moment is roughly estimated as 
$M_{\rm z}^{\rm AF} \approx \Gamma_{\rm zc} \mu_{\rm B} j abc/|e|v_{\rm F}$ per unit cell, where $v_{\rm F}$ [cm/s] 
is the typical Fermi velocity. 
For $ \Gamma_{\rm zc} =10$, $a=b=c=10^{-7}$ cm, and $v_{\rm F} = 10^{7}$ cm/s, an electric current density 
$j=1$ [A/cm$^2$] induces an antiferromagnetic moment $M_{\rm z}^{\rm AF} \sim 10^{-8} \mu_{\rm B}$. 
Although the DMEC of zigzag chains, $ \Gamma_{\rm zc}  \sim 10$, is much larger than that of 
noncentrosymmetric metals, $\Gamma_{\rm zc}  \sim 10^{-1}$~\cite{Comment2}, 
the current-induced magnetic moment is still small. 
In the next section, we show that the current-induced antiferromagnetic moment 
is significantly enhanced in Kondo systems.

\section{Kondo Systems}

We have studied the antiferromagnetic spin polarization of conducting electrons 
induced by electric current. Although the typical moment was small, 
the spin polarization is significantly enhanced in Kondo systems, 
where conduction electrons are coupled with localized spins via Kondo exchange coupling, 
as realized in various d-electron and f-electron systems.

It is reasonably expected that the current-induced antiferromagnetic spin polarization 
of conduction electrons leads to a substantial polarization of localized spins. 
We here consider temperatures above the Kondo temperature and ignore the screening of localized spins 
due to the Kondo effect. Thus, we obtain the Landau free energy 
%$
\begin{eqnarray}
&& \hspace{-4mm}
F = \frac{1}{2\chi_{\rm c}^{\rm AF}} \left[ \left(M_{\rm c}^{\rm AF}\right)^2 
- 2 M_{\rm c0}^{\rm AF} M_{\rm c}^{\rm AF} \right] 
\nonumber \\ && \hspace{2mm}
+ \frac{J_{\rm K}}{gg_{\rm J}\mu_{\rm B}^2} M_{\rm c}^{\rm AF} M_{\rm f}^{\rm AF} 
+ \frac{1}{2\chi_{\rm f}^{\rm AF}} \left(M_{\rm f}^{\rm AF}\right)^2,   
\label{Landau}
\end{eqnarray} 
where $M_{\rm c}^{\rm AF} = g \mu_{\rm B} \langle s_{z}^{\rm c} \rangle$ and 
$M_{\rm f}^{\rm AF} = g_{\rm J} \mu_{\rm B} \langle S_{z}^{\rm f} \rangle$ are 
the antiferromagnetic moments of conduction electrons and localized spins, respectively. 
We have taken into account the Lande $g$-factor of localized spin, $g_{\rm J}$.  
We denote the current-induced antiferromagnetic moment calculated in Sects.~3-5 as 
$M_{\rm c0}^{\rm AF} = - \frac{g\mu_{\rm B}}{2|e|D} \Gamma_{\rm zc} J_{\rm c}$. 
The Kondo exchange coupling $J_{\rm K} \sSS_{i}^{\rm c} \cdot \SS_{\rm i}^{\rm f}$ is taken into account 
in the second term of Eq.~(\ref{Landau}), and 
$\chi_{\rm c}^{\rm AF}$ ($\chi_{\rm f}^{\rm AF}$) is the antiferromagnetic spin susceptibility of 
conduction electrons (localized spins). 
It is easy to minimize the Landau free energy for the following antiferromagnetic moment   
\begin{eqnarray}
\label{M_c}
&& \hspace{-4mm}
M_{\rm c}^{\rm AF} = \beta M_{\rm c0}^{\rm AF}, 
\\ && \hspace{-4mm}
M_{\rm f}^{\rm AF} = - \beta J_{\rm K} \frac{\chi_{\rm f}^{\rm AF}}{gg_{\rm J}\mu_{\rm B}^2} M_{\rm c0}^{\rm AF},  
%= -\eta M_{\rm c0}^{\rm AF}, 
\label{M_f}
\end{eqnarray}
where 
\begin{eqnarray}
&& \hspace{-4mm}
\beta = 
\left[1-J_{\rm K}^2 \chi_{\rm c}^{\rm AF}\chi_{\rm f}^{\rm AF}/(gg_{\rm J}\mu_{\rm B}^2)^2\right]^{-1}. 
\label{beta}
\end{eqnarray}
In this way, the antiferromagnetic moment of localized spins is induced by the magnetoelectric effect 
of conduction electrons through Kondo exchange coupling. 
A large enhancement factor, $\eta \equiv \beta J_{\rm K} \chi_{\rm f}^{\rm AF}/gg_{\rm J}\mu_{\rm B}^2$, 
is obtained at low temperatures because of the large antiferromagnetic spin susceptibility of 
localized spins. 
When we consider nearly isolated ions with the spin $J$, the spin susceptibility is calculated as 
$\chi_{\rm f}^{\rm AF} = g_{\rm J}^2 \mu_{\rm B}^2 J(J+1)/3T$. 
By taking the effect of the antiferromagnetic exchange coupling of localized spins  
as well as the factor $\beta$ into account, we obtain the enhancement factor  
\begin{eqnarray}
\label{eta}
&& \hspace{-10mm}
\eta = \beta J_{\rm K} \frac{\chi_{\rm f}^{\rm AF}}{gg_{\rm J}\mu_{\rm B}^2} 
= J(J+1) \frac{g_{\rm J}}{g} \frac{J_{\rm K}}{3(T-T_{\rm N})}, 
\end{eqnarray}
where $T_{\rm N}$ is the \Neel temperature. 
Thus, a large antiferromagnetic polarization of localized spins is induced 
at low temperatures and/or near the \Neel temperature.

It is interesting to look for the magnetoelectric effect and inverse magnetoelectric effect 
in LnM$_2$Al$_{10}$ compounds, whose crystal structure is indeed 3D coupled zigzag chains~\cite{Thiede}. 
Recent experimental studies of a series of LnM$_2$Al$_{10}$ compounds have shown that 
the antiferromagnetic order occurs in most of these 
compounds~\cite{Thiede,Reehuis,Khalyavin,Tanida_Nd,Muro_Nd-Gd,H.Kato}. 
A variety of inter-zigzag-chain magnetic structures appear depending on the Ln and M ions. However, 
the intra-zigzag-chain magnetic structure is universal; the magnetic moment is staggered in the zigzag 
chains, as shown in Fig.~1. Therefore, it is indicated that the antiferromagnetic correlation in zigzag 
chains enhances the magnetoelectric effect above the \Neel temperature. 
Then, the current-induced antiferromagnetic order should occur; the electric current along 
the {\it a}- or {\it c}-axis increases the \Neel temperature, 
but that along the {\it b}-axis does not.

The enhancement factor of the magnetoelectric effect, $\eta$, is particularly large in LnM$_2$Al$_{10}$ 
compounds because of their electric structure. First, the local violation of the inversion symmetry 
allows for the intrasite hybridization of conducting d-electrons and localized f-electrons 
on Lanthanoid ions. This ``parity mixing'' of local orbitals gives rise to not only a large  
Rashba spin-orbit coupling but also a large Kondo exchange coupling $J_{\rm K}$~\cite{Kunimori}, 
and thus, enhances the magnetoelectric effect. 
Second, localized f-electrons have a large spin for Ln=Nd, Gd, Tb, Dy, Ho, or Er ions. 
Indeed, the magnetic properties of NdFe$_2$Al$_{10}$, NdRu$_2$Al$_{10}$, GdRu$_2$Al$_{10}$, TbRu$_2$Al$_{10}$, 
DyFe$_2$Al$_{10}$, HoRu$_2$Al$_{10}$, and ErRu$_2$Al$_{10}$  have been 
extensively studied.~\cite{Thiede,Reehuis,Khalyavin,Tanida_Nd,Muro_Nd-Gd,H.Kato} 
For instance, $J=9/2$ for Nd$^{3+}$ ions, $J=6$ for Tb$^{3+}$ ions, and $J=8$ for Ho$^{3+}$ ions 
lead to a large enhancement factor $\eta$. 
Third, for transition-metal ions, {\it e.g.} M$=$Ru and Os, a large LS coupling of Ru4d and Os5d electrons 
increases the staggered antisymmetric spin-orbit coupling of conduction electrons~\cite{YanaseCePt3Si},
which is the cause of the magnetoelectric effect.

As an example, we estimate the enhancement factor $\eta$ of NdRu$_2$Al$_{10}$. 
Because the crystal electric field of Nd$^{3+}$ ions is negligible~\cite{Tanida_Nd}, 
our result for the nearly isolated ions, that is, Eq.~(\ref{eta}), is justified. 
The g-factor of Nd$^{3+}$ ions is $g_{\rm J} = 8/11$ and the \Neel temperature is 
$T_{\rm N}=2.4$ K~\cite{Tanida_Nd}. When we assume $J_{\rm K} =400$ K and $T =2.5$ K, 
the enhancement factor is estimated as $\eta = 1.2 \times 10^4$. 
Thus, we expect that a moderate antiferromagnetic moment, $M_{\rm f}^{\rm AF} \sim 10^{-4} \mu_{\rm B}$, 
is induced by the electric current density $j = 1$ [A/cm$^2$]. 
Similar estimations show a larger magnetic moment for Tb, Dy, and Er compounds. 
These current-induced magnetic moments are not large, 
but the experimental observation is feasible.

We have considered the high-temperature region above the Kondo temperature. 
At low temperatures, localized spins are screened by conduction electrons, and
an itinerant heavy-fermion state is formed. 
The magnetoelectric effect in the noncentrosymmetric heavy-fermion state has been investigated 
by Fujimoto~\cite{Springer,Fujimoto2007,Fujimoto_review} on the basis of the Fermi liquid theory. 
He showed that the magnetoelectric effect is enhanced by the mass enhancement factor $m^*/m$. 
His result is reproduced in our calculation by replacing the temperature $T$ 
with the Kondo temperature $T_{\rm K}$. In other words, the magnetoelectric effect is cut off 
at low temperatures by the formation of heavy-fermion states. 
Fortunately, this cutoff is avoided for localized spins with 
a large spin $J$, because the Kondo temperature is exponentially small~\cite{Okada,Yamada1984,Yanase1997}. 
%\begin{eqnarray}
%&& \hspace{-8mm}
%T_{\rm K} = D \exp\left[-\frac{2J+1}{\rho J_{\rm K}}\right].
%\end{eqnarray} 
Indeed, the Kondo effect is not observed even at low temperatures of $T \sim 1$ K  
in Ln=Nd, Gd, Tb, Dy, Ho, and Er compounds. 
It is desirable to study these compounds for the experimental observation of 
the magnetoelectric effect in metals.

\section{Summary and Discussion}

In this paper, we investigated the magnetoelectric effect in locally noncentrosymmetric metals. 
The ``antiferromagnetic moment'' in the unit cell is polarized by electric current 
through a staggered antisymmetric spin-orbit coupling, in analogy with the current-induced 
spin polarization in noncentrosymmetric metals. 
Although the cancellation of spin-split bands significantly suppresses the 
magnetoelectric effect of noncentrosymmetric metals, this cancellation does not occur in 
locally noncentrosymmetric metals. We showed that such a situation is realized in 3D 
coupled zigzag chains. 
Therefore, the magnetoelectric effect in 3D zigzag chains can be much larger than that in  
noncentrosymmetric metals.

Although the induced antiferromagnetic moment of 3D coupled zigzag chains is still small, 
it is enhanced by Kondo exchange coupling with localized spins. 
A particularly large magnetoelectric effect will appear in Kondo systems having 
a large local spin $J$, a large Kondo exchange coupling $J_{\rm K}$, and a moderate antisymmetric 
spin-orbit coupling of conduction electrons. 
We proposed that LnM$_2$Al$_{10}$ compounds, such as NdRu$_2$Al$_{10}$ and TbRu$_2$Al$_{10}$, 
are candidate materials for the experimental observation of the current-induced antiferromagnetic order. 
We roughly estimated the current-induced magnetic moment  
as $M_{\rm f}^{\rm AF} \sim 10^{-4} \mu_{\rm B}$ for an electric current density $j = 1$ [A/cm$^2$].

At low temperatures, the antiferromagnetic order occurs in LnM$_2$Al$_{10}$ compounds 
in the equilibrium state. 
The staggered magnetic moment universally appears in zigzag chains, but various inter-zigzag-chain 
magnetic structures have been observed depending on lanthanoid and transition-metal ions.  
It is expected that the electric current not only increases the \Neel temperature, but also 
changes the magnetic structure. The uniform magnetic state between the network of zigzag chains is favored. 
These phenomena are induced by an effective staggered magnetic field generated by 
the magnetoelectric effect. 
Although it is difficult to externally apply a staggered magnetic field to solids, 
spin-orbit coupling in locally noncentrosymmetric metals changes the electric field to 
a staggered magnetic field.  
It is desirable to experimentally study these magnetoelectric effects as well as the 
inverse magnetoelectric effect of locally noncentrosymmetric metals, 
which have not been uncovered yet.

\section*{Acknowledgements}

The authors are grateful to S. Fujimoto, H. Harima, H. Kusunose, M. Sera, and H. Tanida 
for fruitful discussions. 
We especially thank T. Arima for discussions on the asymmetric band structure in the 
antiferromagnetic state.  
This work was supported by Grants-in-Aid for Scientific Research 
on Innovative Areas ``Heavy Electrons" (No. 23102709) and ``Topological Quantum Phenomena'' 
(No. 25103711) from MEXT Japan, 
and by a Grant-in-Aid for Young Scientists (No. 24740230) from JSPS. 
Part of the numerical computation in this work was carried out 
at the Yukawa Institute Computer Facility.

\appendix

\section{Absence of Spontaneous Electric Current in Antiferromagnetic State}

We here provide a proof for the absence of spontaneous electric current in the 
``antiferromagnetic state''. 
Although the band structure is asymmetric as shown in Fig.~2, the electric current vanishes 
in the equilibrium state.

The electric current in the equilibrium state is calculated as 
\begin{eqnarray}
\label{current}
&& \hspace*{-8mm}  
\J = \langle \hat{\J} \rangle = {\rm Tr} \hat{\J} e^{-\beta \hat{H}}/Z, 
\end{eqnarray}
with the use of the current operator $\hat{\J}$ and the Hamiltonian $\hat{H}$. 
When the Hamiltonian is expressed in single-particle form 
as in Eq.~(\ref{H_k}), 
\begin{eqnarray}
H= \int {\rm d}\k \hat{C}^{\dag}_{\k} \hat{H}(\k) \hat{C}_{\k},  
\end{eqnarray}
the electric current is obtained as 
\begin{eqnarray}
\label{current-2}
&& \hspace*{-8mm}  
\J = \int {\rm d}\k  {\rm Tr} \hat{\J}(\k) e^{-\beta \hat{H}(\k)} /Z_{\k}, 
\end{eqnarray}
where 
\begin{eqnarray}
\label{current-3}
&& \hspace*{-8mm}  
\hat{\J}(\k) \equiv \frac{\rm d}{{\rm d} \k} \hat{H}(\k). 
\end{eqnarray}
Diagonalizing the Hamiltonian via the unitary transformation 
$\hat{H}^{\rm d}(\k) \equiv U^{\dag}(\k) \hat{H}(\k) U(\k) = (E_{i}(\k) \delta_{ij})$, we obtain 
\begin{eqnarray}
\label{current-4}
&& \hspace*{-8mm}  
\J =  \int {\rm d}\k \sum_{l=1}^{n} \J_{l}^{\rm d}(\k) f[E_{l}(\k)], 
\end{eqnarray}
with $\J_{l}^{\rm d}(\k)$ being the $l$-th diagonal component of 
$\hat{\J}^{\rm d}(\k) \equiv U^{\dag}(\k) \hat{\J}(\k) U(\k)$ and $f(x)$ being the Fermi distribution 
function. 
Using the identity for the unitary matrix 
$\left(\frac{\rm d}{{\rm d} \k} \hat{U}^{\dag}(\k)\right) \hat{U}(\k) + \hat{U}^{\dag}(\k)
\left(\frac{\rm d}{{\rm d} \k} \hat{U}(\k)\right) = 0$, $\hat{\J}^{\rm d}(\k)$ is expressed as 
\begin{eqnarray}
\label{current-5}
&& \hspace*{-12mm}  
\hat{\J}^{\rm d}(\k) = \frac{\rm d}{{\rm d} \k} \hat{H}^{\rm d}(\k) 
+ \left[ 
U^{\dag}(\k) \frac{\rm d}{{\rm d} \k} U(\k),  \hat{H}^{\rm d}(\k)
\right]. 
\end{eqnarray}
Because the second term in Eq.~(\ref{current-5}) is anti-Hermite, 
the diagonal component is obtained as $\J_{l}^{\rm d}(\k) = \frac{\rm d}{{\rm d} \k} E_{l}(\k)$. 
Thus, we find that the spontaneous electric current vanishes: 
\begin{eqnarray}
\label{current-6}
&& \hspace*{-8mm}  
J_i =  \sum_{l=1}^{n} \int {\rm d}\k \left[ \frac{\rm d}{{\rm d} k_i} E_{l}(\k) \right] f[E_{l}(\k)] 
\\ && \hspace*{-5mm}
= \sum_{l=1}^{n} \int {\rm d}\k \frac{\rm d}{{\rm d} k_i} F[E_{l}(\k)] 
\\ && \hspace*{-5mm}
= \sum_{l=1}^{n} \int_{\rm BZB} {\rm d}\k F[E_{l}(\k)] \left( \Ca_i \cdot \Cn \right)
= 0.  
\end{eqnarray}
The last equality is obtained using the periodicity of $E_{l}(\k) = E_{l}(\k+\K)$ 
for the reciprocal lattice vector $\K$. 
The integral $\int_{\rm BZB} {\rm d}\k$ is taken at the Brillouin zone boundary, 
and $\Cn$ is a unit vector normal to the Brillouin zone boundary. 
We denoted the $i$-th fundamental vector $\Ca_i$; $F(x)$ is an indefinite integral of $f(x)$.

\bibliographystyle{jpsj}
\bibliography{66119}% Produces the bibliography via BibTeX.

\begin{thebibliography}{10}
\bibitem{Comment1}
This finding is in contrast to the case of even locally 
centrosymmetric crystals in which the non-collinear magnetic order is required for the violation 
of the inversion symmetry~\cite{Katsura}.
\bibitem{Comment2}
For instance, we obtain $\Gamma_{\rm zc}  \sim 10^{-1}$ for $t_1=0$ and 
$t_2 = t_3 = t_4 = 1$. For these parameters, a zigzag chain is regarded as 
a pair of non-centrosymmetric 1D chains.
\end{thebibliography}

\end{document}